\documentclass[aps,pra,english,showpacs,twocolumn,groupedaddress,superscriptaddress]{revtex4}
\usepackage[english]{babel}
\usepackage[ansinew]{inputenc}
\usepackage{amsfonts}
\usepackage{latexsym}
\usepackage{amsmath}
\usepackage{graphicx}
\usepackage{bm}
\usepackage{natbib}

\begin{document}

\title{Fidelities for transformations of unknown quantum states}

\author{Lars Bojer Madsen}
\affiliation{Department of Physics and Astronomy,
  University of  Aarhus, 8000 {\AA}rhus C, Denmark}
\author{Klaus M{\o}lmer}
\affiliation{Department of Physics and Astronomy,
  University of  Aarhus, 8000 {\AA}rhus C, Denmark}
\affiliation{QUANTOP - Danish National Research Foundation Center
for
  Quantum Optics, Department of Physics and Astronomy, University of Aarhus,
  8000 {\AA}rhus C, Denmark}

\date{\today}

\begin{abstract}
We present a general theoretical formalism to compute the fidelity of  transformations of unknown
quantum states, and we apply our theory to Gaussian transformations of continuous variable quantum
systems. For the case of a Gaussian distribution of displaced coherent states, the theory is readily
tractable by a covariance matrix formalism, and a wider class of states, exemplified by Fock states,
can be treated efficiently by the Wigner function formalism. Given the distribution of input states,
the optimum feed back gain is identified, and analytical results for the fidelities are presented for
recently implemented teleportation and memory storage protocols for continuous variables,
\end{abstract}

%03.67.-a Quantum information
%03.67.Hk Quantum communication
%03.67.Mn Entanglement production, characterization and manipulation
%(see also 03.65.Ud Entanglement and quantum nonlocality; for
%entanglement in Bose-Einstein condensates, see 03.75.Gg)
%03.65.Ud Entanglement and quantum nonlocality (e.g. EPR paradox,
%Bell's inequalities, GHZ states, etc.) (for entanglement production
%in quantum information, see 03.67.Mn; for entanglement in
%Bose-Einstein condensates, see 03.75.Gg)

\maketitle

\section{Introduction}

In a generic scenario for quantum state transformations, a protocol is applied to an unknown input
quantum state taken from a family of states with a certain probability distribution. The quality of
the protocol is quantified by a fidelity measure, which in a natural sense extracts the average
overlap between the state obtained through use of the protocol and the state expected under ideal
circumstances.

In laboratory experiments, one may delegate the handling of the initial random state preparation and
the final examination of the output to an independent person or device, Victor, (and appeal that
Victor is not leaking information to the other experimenters). Theoretical physics works differently
in the sense that what is specified at one point in the theory is specified throughout, and quantum
physics adds the further aspect that the quantum state is a state describing our knowledge about the
system, influenced by any knowledge that comes to our mind be it in the form of measurement outcomes
or information about the preparation procedure for the system - we cannot know something about a
system and at the same time describe it by a state vector or density operator that is independent of
this knowledge.

It is the purpose of this paper to present a practically useful theory to determine fidelities without
leaving doubts about the correct handling of what is known and what is not known about the input
states. In Sec. II, we introduce a theoretical formalism which represents Victor both at the stage of
preparation of an unknown input state and at the examination of the output. In Sec. III, we consider
the case of continuous variable quantum systems, and we show that a hybrid quantum and classical
Wigner function can be applied in an explicit calculation of the fidelity. In Sec. IV, we consider
quantum state teleportation by the Braunstein and Kimble protocol \cite{Braunstein98}, and we identify
the optimum operation given the input state distribution, and we obtain analytical results for the
teleportation fidelity. In Sec. V, we consider a recently implemented quantum memory protocol
\cite{Julsgaard04}, and we also here identify the optimum operation and fidelities. In Sec. VI, we
consider teleportation of non-Gaussian states as exemplified by a distribution of Fock states. Section
VII concludes the paper.

\section{General transformation of an unknown quantum state}

Consider a family of states $\{|\Psi^{in}(\lambda)\rangle\}$, parameterized by a stochastic variable
$\lambda$ and used as input to a certain quantum protocol according to the probability distribution
$P(\lambda)$. For convenience, we assume the different states be obtained from one reference state
$|\Psi_0\rangle$ by means of a family of unitary operators, $|\Psi^{in}(\lambda)\rangle=U_\lambda
|\Psi_0\rangle$. The randomness of $\lambda$ is now handled by introducing an auxiliary (fictitious)
physical system with no free evolution and with an initial mixed state, $\sum_\lambda
P(\lambda)|\lambda\rangle \langle\lambda|$, where we assume that the quantum states $|\lambda\rangle$
are orthonormal. In our theoretical modelling, we let the auxiliary system interact in a
Quantum-Non-Demolition (QND) manner with our physical system prepared in $|\Psi_0\rangle$,
$H=\sum_\lambda |\lambda\rangle\langle\lambda|\otimes H_\lambda$, which we assume will lead, after a
suitable interaction time, to the following correlated state,
\begin{eqnarray}
\label{eq:mixture} \rho =  \sum_\lambda P(\lambda)|\lambda\rangle
\langle\lambda|\otimes|\Psi^{in}(\lambda)\rangle\ \langle\Psi^{in}(\lambda)|.
\end{eqnarray}

The partial trace over the auxiliary $\lambda$-degrees of freedom produces a density operator
$\sum_\lambda P(\lambda)|\Psi^{in}(\lambda)\rangle\ \langle\Psi^{in}(\lambda)|$ describing the mixture
of input states  to the protocol with the appropriate probabilities, but we note that the fidelity of
a protocol is not a measure of how well a mixed state is transformed into its image by the ideal
protocol, but a measure of how well each member of the mixture transforms. It should also be noted
that Eq.(\ref{eq:mixture}) identifies specific individual pure state components in the input, whereas
the reduced density matrix $\sum_\lambda P(\lambda)|\Psi^{in}(\lambda)\rangle\
\langle\Psi^{in}(\lambda)|$ does not have a unique unravelling in terms of pure states. In
(\ref{eq:mixture}) we have retained the variable $\lambda$ in the auxiliary system, which enables us
to study the fidelity at the pure state level, without selecting a specific pure state input to the
protocol.

A physical transformation of a quantum state must be completely positive and preserve the
normalization of the density operator, and it can be written most generally in the Kraus form, $\rho
\rightarrow \sum_s E_s\rho E_s^\dagger$, where the $E_s$ operators can be any set of operators that
fulfils $\sum_s E_s^\dagger E_s = 1$. Important examples include (i) unitary evolution, where there is
only one unitary operator, $E_s=U $, (ii) open system dynamics following a Lindblad form master
equation with  jump and no-jump operators $E_s$, (iii) von Neumann measurements of a hermitian system
observable with orthogonal projections $E_s$, and (iv) more general measurement positive operator
valued measure (POVM) scenarios.

Without loss of generality we assume that the desired protocol takes our quantum state to a final
state on a similar Hilbert space (same dimensionality), so that each input state
$|\Psi^{in}(\lambda)\rangle$ is ideally transformed by a unitary operation $V$ into
$|\Psi^{out}(\lambda)\rangle=V U_\lambda |\Psi_0\rangle$. We assume that physical interaction and
measurements take place only on our quantum system of interest and on possibly added quantum
systems.The variable $\lambda$ is not made subject to interaction or direct observation, and we hence
write the state after application of the protocol, tracing over possible further unobserved quantum
degrees of freedom as
\begin{equation}
\label{eq:Kraus} \rho=\sum_\lambda P(\lambda)|\lambda\rangle \langle\lambda|\otimes\left(\sum_s
E_s|\Psi^{in}(\lambda)\rangle\langle\Psi^{in}(\lambda)|E_s^\dagger \right).
\end{equation}
The sum over $E_s$ terms may represent the result of measurements on the system and it may also
incorporate a unitary feedback $U^{\textrm{feedback}}_s$ on the system, conditioned on the outcome $s$
of the measurement. ($E_s \rightarrow U^{\textrm{feedback}}_s E_s$ also fulfil the required property
$\sum_s (U^{\textrm{feedback}}_s E_s)^\dagger U^{\textrm{feedback}}_s E_s)=1$ of the Kraus form).

Equation (\ref{eq:Kraus}) is very illustrative. The sum over the different Kraus operators corresponds
to averaging over the outcomes of measurements and potential feedback on the system. It shows how each
input component transforms into a mixed state $\rho^{out}(\lambda)=\sum_s
E_s|\Psi^{in}(\lambda)\rangle\langle\Psi^{in}(\lambda)|E_s^\dagger$. If we accept the output for all
such measurement results, we must carry out this average, and we use the state in Eq.(\ref{eq:Kraus})
to compute the fidelity of the protocol. To check if $\rho^{out}(\lambda)$ equals the desired state
$|\Psi^{out}(\lambda)\rangle=V U_\lambda |\Psi^{in}(\lambda)\rangle$ we apply the unitary
$\sum_\lambda |\lambda\rangle \langle\lambda|\otimes(VU_\lambda)^{-1}$ on (\ref{eq:Kraus}), and check
if the quantum system is now in the initial reference state $|\Psi_0\rangle$. Rather than verifying
that the output equals the ideally transformed input state, we check if the inverse of the tranform on
the output agrees with the fixed reference input state. Ideally, this agreement should be obtained for
all $\lambda$-components of the system and we thus perform the partial trace over the
$\lambda$-degrees of freedom and compare the final state system density matrix  with the pure state
$|\Psi_0\rangle$:
\begin{equation}
F=\sum_\lambda P(\lambda) \langle \Psi_0|U_\lambda^\dagger V^\dagger \rho^{out}(\lambda) V
U_\lambda|\Psi_0\rangle.
\end{equation}
In this scheme we compute the average value of the overlap without having to specify which initial
state is applied.

We observe that this result can also be written
\begin{equation}\label{eq:fidsimpl}
F=\sum_\lambda P(\lambda) \langle \Psi^{out}(\lambda)|\rho^{out}(\lambda)|\Psi^{out}(\lambda)\rangle,
\end{equation}
where $|\Psi^{out}(\lambda)\rangle=V|\Psi^{in}(\lambda)\rangle$ is the desired output state, and
despite our concerns in the Introduction about a consistent treatment of unknown input states, the
fidelity \textit{is} simply the fidelities obtained for each input state averaged over the input state
distribution. A measurement part of the protocol may yield some information about the
$\lambda$-variable and hence change the probability distribution $P(\lambda )$, but when we average
over the outcomes $s$, we return to the original distribution. The fact that a feedback may be applied
to the system conditioned on the measurement affects only the fidelity through the form of the $E_s$
operators. We note that Eq.(\ref{eq:fidsimpl}) could give the erroneous impression that the
$P(\lambda)$ distribution only enters via the explicit weighted sum. This is not the case; as we shall
see below, to yield the highest possible fidelity the optimum feedback, i.e., the operators
$U^{\textrm{feedback}}_s$ should be chosen in a manner that depends on the distribution $P(\lambda)$.

If the tranformation can be applied with a non-unit success probability, i.e., if the output state is
only accepted conditioned on a specific outcome or set of outcomes $\{s'\}$ of the measurement on the
system, we must go back to the joint state (\ref{eq:Kraus}), restrict the sum to only these values and
renormalize the state. (The trace of the un-normalized state is precisely the probability of
acceptance). The resulting state is now a weighted sum of density operator terms with non-unit trace
$q_\lambda = \textrm{Tr}(\sum_{s'}
E_{s'}|\Psi^{in}(\lambda)\rangle\langle\Psi^{in}(\lambda)|E_{s'}^\dagger)$. We can renormalize the
density operators with $1/q_\lambda$ and multiply the same $q_\lambda$ factors on $P(\lambda)$ which
represents then the updated probability distribution of the input states conditioned on the
measurement result. Our fidelity calculation proceeds as above with the inverse transformations and
the final comparison with the initial reference state, and in this case, the fidelity is again given
by the state-to-state transformation fidelities but now weighted by both their initial state
probability and their individual probabilities $q_\lambda$ for the acceptable measurement outcome.

\section{Application to continuous variable systems}

Quantum information protocols with continuous variable systems have been the focus of intense research
since it was suggested \cite{Braunstein98} and demonstrated \cite{Furusawa98} that existing squeezed
light sources, beam splitters and photodetectors suffice to enable quantum state teleportation of
light. The collective atomic population of different internal states in a macroscopic gas sample also
provides effectively continuous degrees of freedom, and efficient atomic entanglement protocols that
make  use of classical light sources and photodetection only were proposed \cite{duan00} and
demonstrated \cite{Julsgaard01}. The work on entangled atomic gasses was followed by theoretical and
experimental work on quantum state transfer between light and matter (a quantum memory for light
\cite{Julsgaard04,Sherson05-qubit}), and ideas for atomic state teleportation \cite{duan00} are
currently being pursued.

These continuous variable systems can be described in terms of canonically conjugate harmonic
oscillator variables $x$ and $p$, and states can be described in terms of Wigner phase space
distribution functions in place of the general density matrix notation of the previous section. We
consider the case where the ensemble of input states is obtained by displacements of the reference
state (the vacuum state in Secs. IV and V) by arguments $x_{\textrm{cl}}$ and $p_{\textrm{cl}}$
according to a probability distribution $P(\lambda=(x_{\textrm{cl}},p_{\textrm{cl}}))$. Such a
displacement of a Wigner function simply amounts to a translation of its argument $W(x,p) \rightarrow
W(x-x_\textrm{cl},p-p_\textrm{cl})$, but as in Sec. II we shall introduce an auxiliary set of QND
variables in the modelling of the input state ensemble. We thus treat the real arguments
$x_{\textrm{cl}}$ and $p_{\textrm{cl}}$ as two independent variables, e.g., momenta for free
particles, or  simply as classical variables in a quantum-classical hybrid Wigner function for the
total system, which is consistent with Heisenberg's uncertainty relation for the quantum degrees of
freedom, but has no such constraints on the classical degrees of freedom. If we take the zero
amplitude coherent state with a Wigner function $W_0(x,p)$ and displace it by the classical arguments
$x_{\text{cl}},p_{\text{cl}}$ according to a classical probability distribution
$P(x_{\text{cl}},p_{\text{cl}})$, the joint Wigner function of the quantum and classical variables
become in analogy with (\ref{eq:mixture}),
\begin{equation}\label{Wigner-in}
W_{\textrm{in}}(x,p,x_{\text{cl}},p_{\text{cl}})=W_0(x-x_{\text{cl}},p-p_{\text{cl}})P(x_{\text{cl}},p_{\text{cl}}).
\end{equation}

Some quantum information protocols make use of additional quantum systems and we shall hence work with
a multi-variable Wigner function for all the quantum systems and classical variables involved in the
protocol. In teleportation, for example, the communication channel is described by a joint Wigner
function of the entangled state of two quantum systems $W_{\textrm{ent}}(x_1,p_1,x_2,p_2)$. The total
Wigner function is thus a function of 8 variables
$W(x,p,x_{\text{cl}},p_{\text{cl}},x_1,p_1,x_2,p_2)$. The linear mode mixing transformations of the
teleportation protocol amount to the application of linear transformations on the variables within the
original distribution function; measurements of a given quantum variable amounts to evaluating the
function with the corresponding argument attaining the measured value (and integrating over the
canonical conjugate variable which is accordingly completely undetermined), and finally a joint
distribution of the output quantum state and the classical variables is obtained. The verification of
the protocol consists in comparing the output state with the desired one (which for teleportation is
the same as the input state) and this is done by displacing the quantum system with the negative of
the classical variables (inverse of Eq.(\ref{Wigner-in})), integrating over the unknown classical
variables and comparing the ensuing quantum state with the reference state $W_0(x,p)$.

In the most general case one has to deal with a multi-variable function, and one has to carry out
integrals of this function with respect to a number of the variables. We shall now turn to examples
where the initial quantum states and the classical distribution function are all Gaussian. This
situation is of practical relevance in the above mentioned experiments and it offers a significant
simplification of the problem. Gaussian states are fully characterized by the mean values and the
covariance matrix for the variables, and quantum state overlap integrals are given explicitly by these
quantities. The present approach to the fidelity problem, involving joint probabilities for classical
and quantum variables, is related to recent applications of the general theory of parameter estimation
and Kalman filtering with quantum systems \cite{Maybeck,stockton}, and in particular to our recent
application of this theory to atomic magnetometry \cite{Moelmer04,Petersen05}. The formal treatment of
the fidelity issue is actually simpler than the magnetometry analysis, and we now present the details
of such a calculation.

\section{Teleportation of an unkown coherent state}

We treat  the case of teleportation of a physical system 3 by use of an entangled pair of systems, 1
and 2. It has been argued, that a general positive map can be viewed as "teleporting a state through a
gate"~\cite{GiedkeCirac}, and hence this operation has both specific and more general interest.

We note that for coherent states with an amplitude of a given absolute value but with a random choice
of complex phases, Ide et al. studied how to optimize the teleportation fidelity by a proper choice of
the strength of the feedback on the output fields~\cite{Ide02}. We will consider states with a
Gaussian distribution of complex amplitudes, and we will also identify an optimum feedback strength.
Fiurasek has applied the covariance matrix formalism, similar in spirit to our work, but rather than
optimizing he assumed a fixed value for the feedback strength, and then he turned to a study of the
effect of further local Gaussian operations ~\cite{Jaromir02}.

\subsection{Covariance matrix method}

We shall be working with Gaussian states, and hence the state is fully characterized by the mean
values $m_i=\langle y_i \rangle$  and the covariance matrix $\gamma_{ij}=2 \textrm{Re}(\langle
(y_i-\langle y_i\rangle)( y_j-\langle y_j\rangle)$ of all quadrature variables $y_i$. For a more
detailed description of the covariance matrix formalism and its practical implementation of linear
transformations and measurement processes, see, e.g., \cite{Madsen05}.

Following~\cite{Braunstein98}, we use the entanglement in the 12-system to teleport an unknown
coherent state of system 3 drawn from an ensemble of states with a Gaussian distribution of the mean
amplitude onto system 1 by performing displacements of system 1 conditioned on the output of joint
measurements on systems 2 and 3. We introduce the  auxiliary classical variables
($x_\text{cl},p_\text{cl})$ with zero mean and variance given by
$v_\text{c}=2\text{Var}(x_\text{cl})=2\text{Var}(p_\text{cl})$. The classical variables represent an
agent Victor who's role is to turn the vacuum input state of variables $(x_3,p_3)$ into a coherent
state by a displacement. Experimentally, one is interested in the case with sizable $v_\text{c}$
(truly unknown input states), but we shall obtain general expressions for arbitrary $v_c$. Note that
since $v_c$ describes classical variables, it is not limited by the Heisenberg uncertainty relation,
and $v_c=0$ corresponds to the case, where the input is the vacuum state with certainty. The
covariance matrix of system 3 and Victors classical variables prior to the displacement is
$\gamma_{3,V}=\text{diag}(1,1,v_\text{c},v_\text{c})$, and the displacement leads to the
transformation $\gamma_{3,V} \rightarrow \gamma_{3,V}' = S_d \gamma_{3,V} S_d^T$, where the matrix
$S_d$ describes the linear mapping $x_3 \rightarrow x_3+x_\text{cl}, p_3 \rightarrow p_3+p_\text{cl},
x_\text{cl} \rightarrow x_\text{cl}, p_\text{cl}\rightarrow p_\text{cl}$. The mean values are also
formally transformed by this mapping, but since the classical distribution and the quantum state are
assumed to have vanishing mean values initially, this is also the case after the action of $S_d$.

After this preparation of a quantum input state correlated with classical stochastic variables as
described by $\gamma_{3,V}'$, we obtain the complete $8\times 8$ covariance matrix
$\gamma=\textrm{blockdiag}(\gamma_{12},\gamma_{3,V}')$, with $\gamma_{12}$ the covariance matrix for
the initially entangled 12 system. The commuting pair of non-local variables $x_-^{(23)}=
(x_2-x_3)/\sqrt{2}$, $p_+^{(23)}= (p_2+p_3)/\sqrt{2}$ is measured. It is useful to transform the
system to the following eight variables: ($x_1,
p_1,x^{(23)}_+,p_+^{(23)},x_-^{(23)},p_-^{(23)},x_\text{cl},p_\text{cl}$) with the covariance matrix
$\gamma \rightarrow \gamma' = T \gamma T^T$  where the block diagonal matrix $T=
\text{blockdiag}(I_2,T^{(23)},I_2)$, $I_2$ the $2 \times 2$ identity matrix (system 1 and the
classical displacements are not affected by this transformation), and $T^{(23)}$ the matrix effecting
the change of basis from system 2 and 3 variables to the joint variables $x_\pm^{(23)},p_\pm^{(23)}$.
A measurement of a single variable from a joint Gaussian distribution results in an updated Gaussian
distribution for the remaining unknown variables. This update is readily accounted for in terms of the
mean values and the covariance matrix of the variables. First, we reorganize the variables in the
order $(x_1, p_1,x_\text{cl},p_\text{cl},x^{(23)}_+,p_+^{(23)},x_-^{(23)},p_-^{(23)})$, so that the
covariance matrix $\gamma'$ is decomposed into $4\times 4$ dimensional matrix blocks:
\begin{equation}
\gamma' =\left(
\begin{array}{cc}
  A & C \\
  C^T & B \\
\end{array}
\right),
\end{equation} where $A$ is the covariance matrix for the unmeasured quantum and classical components, $B$ is the covariance matrix for
the measured variables, and $C$ and $C^T$ describe the correlations between the measured an unmeasured
variables. The effect of the measurement on a subsystem on the covariance matrix of the remaining,
unmeasured variables is given by the update formula \cite{EisertPlenio,Madsen05}
\begin{equation} A \rightarrow A' = A -C (\pi B \pi )^- C^T,
\end{equation} where $\pi = \text{diag}(0,1,1,0)$ with unity at the entrances of $p_+^{(23)},
x_-^{(23)}$ and $(\dots)^-$ denotes the Moore-Penrose pseudoinverse.

The measurement outcome affects the mean values: conditioned on a positive readout $\xi_{23}$ in the
measurement of $x^{(23)}_-$ our knowledge about $x_3$ and hence of $x_{\text{cl}}$ is biased towards
negative values and $x_2$ is biased towards positive values (and hence $x_1$ is biased towards
positive values). Precisely how much, is determined by the variances initially ascribed to these
variables, and we have the following formula~\cite{EisertPlenio,Madsen05} for the vector of mean
values,  $\textbf{m}=\langle (x_1,p_1,x_\text{cl}, p_\text{cl})^T\rangle$: \begin{equation}\textbf{m}
\rightarrow C (\pi B \pi )^-(\cdot,\eta_{23},\xi_{23},\cdot)^T, \end{equation} where the dots replace
unmeasured quantities, which do not need to be specified due to the zeros in the projector $\pi$, and
$\eta_{23}$ and $\xi_{23}$ denote the random outcome of the measurements of $p_+^{(23)}$ and
$x_-^{(23)}$. In Sec IV B, we shall give the expressions for the change of mean values conditioned on
the random measurement outcome. It is optimal to apply a feedback that brings the mean value of the
quantum variables in exact agreement with the mean value of the classical displacement terms. This is
so, because the covariance matrix does not depend on the measurement outcome, hence at the end of the
calculation we shall compare the vacuum Gaussian state with either a single Gaussian state with
vanishing mean or with a distribution of Gaussian states with the same covariance matrix but with
different displacements with respect to the desired state.

The feedback, just described is part of the teleportation protocol. As part of our verification or
quality assessment of the protocol, we displace the final quantum state by the negative of the
classical parameters and compare the outcome, known to have vanishing mean amplitudes, with the vacuum
state. Correspondingly, we apply the inverse of the classical displacement $S_d^{-1}$ on the quantum
and classical variables $x_1,p_1,x_\textrm{cl},p_\textrm{cl}$ and obtain their resulting covariance
matrix, $V=S_d^{-1} A' (S^{-1})^T$. The $2\times 2$ block $\gamma_{\textrm{out}}=V(1:2,1:2)$
describing the covariances of the quantum variables is the covariance matrix for the quantum system,
when the unknown classical displacements are integrated out,  and it should ideally be the identity
matrix describing the vacuum state.

The fidelity of the protocol is the overlap of the Wigner functions. For a single mode state the
Wigner function is given by $W=1/(\pi\sqrt{\text{det}\gamma})\exp(-\chi^T \gamma^{-1} \chi)$ with
$\chi^T=(x,p)$. In terms of Wigner functions, the average fidelity is defined as $F=2 \pi
\int_{-\infty}^\infty dx \int_{-\infty}^\infty dp W_\text{in} (x,p)W_\text{out}(x,p)$. The vacuum
state is Gaussian with a covariance matrix equal to the identity $I_{2}$, and the integrant is thus a
Gaussian function $\propto \exp(-\chi^T \gamma_\text{res}^{-1} \chi)$ with
$\gamma_\text{res}=(\gamma_\text{out}^{-1}+I_{2})^{-1}$ so that the integral follows directly from the
standard expressions for Gaussian normalization integrals,
\begin{equation}
\label{eq:F1-tel} F= 2{\sqrt \frac{ \text{det}(\gamma_\text{res})} {\text{det}(\gamma_\text{out})}}.
\end{equation}

\subsection{Results}

For simplicity we consider the symmetric case where the joint covariance matrix of the variables
$(x_1,p_1,x_2,p_2)$ for systems 1 and 2 is given by
\begin{eqnarray}
\label{eq:cov1-tel} \gamma_{12} = \left(
  \begin{array}{cccc}
    n & 0 & k & 0 \\
    0 & n & 0 & -k \\
    k & 0 & n & 0 \\
    0 & -k & 0 & n \\
  \end{array}
\right),
\end{eqnarray}
where $n$ describes twice the variance of the variables of system 1 and 2, and where $k$ describe the
correlations between the systems. The collective variables $(x_1\pm x_2)$ and $(p_1\mp p_2)$ have the
variances $(n\pm k)$, and the Heisenberg uncertainty relation implies that $n^2 - k^2 \ge 1$.
Realizations of such a bipartite entangled state include the atom-light setup~\cite{Julsgaard01} and
the EPR-light source channel~\cite{Furusawa98}. The matrix operations just described are
straightforward, and we readily obtain analytical results at all steps of the calculation. The
measurement process yields random outcomes $\xi_{23},\eta_{23}$, and inserting the initial covariance
matrices described above and carrying out the matrix operations, we obtain the conditioned mean values
$\langle x_1\rangle=\frac{k}{1+n+v_c}(\sqrt2 \xi_{23})$ and $\langle
x_\text{cl}\rangle=\frac{-v_c}{1+n+v_c}(\sqrt2 \xi_{23})$, and similar expression for
$p_1,p_{\textrm{cl}}$ in terms of the measured quantity $\eta_{23}$. At this point in the
teleportation protocol, the aim is to have a state with $\langle x_1\rangle = \langle
x_\text{cl}\rangle, \langle p_1\rangle = \langle p_\text{cl}\rangle$, and this is obtained by applying
a feed-back on system 1, in form of a displacement for both the $x_1$ and $p_1$ variables:
\begin{eqnarray}\label{eq:feedbacks}
x_1 \rightarrow x_1 - \frac{k+v_c}{1+n+v_c}(\sqrt{2}\xi_{23}),\nonumber \\
p_1 \rightarrow p_1 + \frac{k+v_c}{1+n+v_c}(\sqrt{2}\eta_{23}).
\end{eqnarray}
In the limit of infinite $v_c$ the feedbacks \eqref{eq:feedbacks} are $\sqrt 2$ times the measured
values themselves, but for states chosen from a finite width distribution, we see that a non-trivial
gain factor
\begin{equation}\label{eq:g-telep}
g=\frac{k+v_c}{1+n+v_c}
\end{equation}
should be applied in the feedback.

The resulting explicit expression for the fidelity reads
\begin{equation}
\label{eq:Fvc-sym} F=\frac {2 \left( 1+n+v_c \right)   }{ \left( 1+2n+n^2-k^{2}+2v_c(1+n-k) \right) }.
\end{equation}

\noindent If the input state is the vacuum state with certainty, $v_c=0$, according to
(\ref{eq:g-telep}), the optimum feedback gain is $g=k/(1+n)$, and we note that for $n^2-k^2=1$, which
characterizes a pure two-mode squeezed state, system 1 is restored in the vacuum state with unit
fidelity.

\noindent In the opposite limit $v_c \rightarrow \infty$, $F$ simplifies to the result
\begin{equation}
F=\frac{1}{1+\Delta},
\end{equation}
where $\Delta=n-k=\textrm{Var}(x_1-x_2)=\textrm{Var}(p_1+p_2)$ is also known as the EPR variance of
systems 1 and 2 \cite{Giedke03}. The fidelity approaches unity when this variance approaches zero
corresponding to a maximally entangled channel.

\noindent Finally, if the quantum channel is in the vacuum state with $n=1$ and $k=0$, the optimum
gain is $g=v_c/(1+n+v_c)$, and our general relation (\ref{eq:Fvc-sym}) reduces to $F =
(2+v_\text{c})/(2+2v_\text{c})=(1+\lambda)/(2+\lambda)$ where $\lambda=2/v_\textrm{c}$, which is
exactly the best result that can be obtained with a classical strategy \cite{Hammerer05}.

\section{Quantum state storage}

Let us now turn to another example: an atomic quantum memory, as demonstrated in a recent
experiment~\cite{Julsgaard04}. In this protocol the aim is to store the quantum state of a light pulse
in the collective spin degrees of freedom of a spin polarized atomic sample. The transverse quantum
degrees of freedom of the collective spin can be effectively described by canonical conjugate
variables (their commutator, the polarized spin component, can be treated as a constant). In the
protocol investigated in \cite{Julsgaard04}, the optical Faraday rotation provides the light-atom
interaction, described by a bilinear interaction Hamiltonian $\propto p_Lp_A$. In the Heisenberg
interaction picture this causes a change in the conjugate variables
\begin{eqnarray}\label{kappa-trans} x_A \rightarrow x_A+\kappa p_L,\nonumber \\ x_L\rightarrow
x_L+\kappa p_A,\end{eqnarray} where $\kappa$ is the dimensionless integrated interaction
strength\cite{Julsgaard04}. The interaction thus encodes the field variable $p_L$ onto the atomic
$x_A$, and by subsequently detecting the $x_L$ component of the field and displacing $p_A$ according
to the measurement result, also this field component is read onto the atoms.

A theoretical analysis of the fidelity of this approach, applied to an unknown coherent state of light
taken from a Gaussian distribution of coherent state amplitudes follows the above discussion of
teleportation. We introduce classical variables $x_{\text{cl}},p_{\text{cl}}$ with variance parameter
$v_c$ and zero mean and quantum variables for the atoms and light in zero mean field coherent initial
states, so that the Wigner function is a function of six variables, and the covariance matrix is
$6\times 6$. We apply the linear transformation between the field variables and classical variables to
initialize the ensemble, and we apply the time evolution due to the atom-light interaction
(\ref{kappa-trans}). These operations cause a mathematical transformation of the covariance matrix,
and all mean values still vanish. The detection of the field component $x_L$ leads to an output value
$\xi$, and conditioned on this output, we obtain the mean values, $\langle p_A\rangle =
\kappa/(1+\kappa^2+v_c)\xi$ and $\langle x_\text{cl} \rangle =v_c/(1+\kappa^2+v_c)\xi$. We wish to
encode $-x_L$ in $p_A$~\cite{Julsgaard04}, and shall hence apply a feedback on the atomic $p_A$
variable $p_A \rightarrow p_A - g\xi$ with the non-trivial (optimal) gain factor
\begin{equation}
\label{eq:gain_memory}
 g=(\kappa+v_c)/(1+\kappa^2+v_c)
\end{equation}

As for teleportation, in the $v_c \rightarrow \infty$ limit the optimum feedback gain is unity, but
for finite width distributions it depends explicitly on the variance $v_c$. The state stored is now
guaranteed to have the same mean amplitudes as the classical variables, and to check if we managed to
store $p_L$ in $x_A$ and $-x_L$ in $p_A$, we follow the procedure from above and displace $x_A$ by
$-p_\text{cl}$ and $p_A$ by $x_\text{cl}$, and compare the resulting Gaussian state covariance matrix
with the vacuum state as in Eq.~\eqref{eq:F1-tel}.

The resulting fidelity is a function of the variance of the classical variables and the coupling
strength $\kappa$:
\begin{widetext}
\begin{equation}
F=2\sqrt{\frac{1+\kappa^2+v_c}{(1+v_c \kappa^2- 2 v_c \kappa + v_c +\kappa^2 + 2 v_c +1)(1 + v_c
\kappa^2 - 2 v_c \kappa + v_c + \kappa^2  +1)}}.
\end{equation}
\end{widetext}

This general expression for the storage fidelity has several interesting limits. First, we observe
that for a completely unknown initial state with $v_c\rightarrow\infty$, the fidelity vanishes unless
$\kappa=1$, in which case one gets the value $F=\sqrt{2/3}\sim 0.8165$, reported in the literature
\cite{Julsgaard04}. In the opposite limit of a known vacuum input, the choice $\kappa=1$ yields
$F=2\sqrt{2}/3\sim 0.9428$ for the storage fidelity. The optimum strategy for finite $v_c$, however,
is to adjust the value of $\kappa$ so as to maximize the fidelity, and this leads to unit storage
fidelity for $\kappa=0,v_c=0$.

In the storage protocol, the field variables are mapped onto the atomic ones, but part of the initial
atomic noise in the $x_A$ variable remains in the atomic system whereas the feedback manages to cancel
the $p_A$ component exactly. It has therefore been suggested to use an initially squeezed atomic
state. This is readily analyzed in our description. We simply take the values ($1/r,r$) with $r$ a
squeezing parameter larger than unity for the initial diagonal elements of the atomic covariance
matrix in the $(x_A,p_A)$ basis and go through all of the above steps again. In this case, we find the
optimum feedback gain factor
\begin{equation}
\label{eq:gain_memory}
 g=(\kappa r+v_c)/(1+\kappa^2 r+v_c)
\end{equation}
The state stored is again guaranteed to have the same mean amplitudes as the classical variables, and
the fidelity of the memory storage is a function of the initial atomic squeezing, the variance of the
classical variables and the coupling strength $\kappa$:
\begin{widetext}
\begin{equation}
F=2\sqrt{\frac{r(1+\kappa^2 r+v_c)}{(r+rv_c \kappa^2- 2  r v_c \kappa + r v_c +\kappa^2 r + 2 v_c
+1)(r + rv_c \kappa^2 - 2 r v_c \kappa + r v_c + \kappa^2 r +1)}}.
\end{equation}
\end{widetext}
This expression yields the storage fidelity, and we again see that if $v_c\rightarrow \infty$, $F$
vanishes unless $\kappa=1$.

\noindent For infinite, or very large, $v_c$ and $\kappa=1$, the fidelity can be expanded in the
squeezing parameter $r\ll v_c$, $F=\sqrt{2r/(2r+1)}$ which yields the already mentioned $\sqrt{2/3}$
for $r=1$ and which approaches unity for large $r$.

\section{Transformation of non-Gaussian states}

For quantum computing it is necessary to be able to store logical states 0 and 1 and their
superposition states, and one may not restrict an analysis to Gaussian states only, but still the
above mentioned protocols may be useful. A Gaussian entangled state may be used to teleport also
non-Gaussian states \cite{Caves04}, and a quantum memory protocol which transforms Gaussian states
into Gaussian states also applies to qubit states, encoded in a two-dimensional subspace of the
continuous variable Hilbert space~\cite{Sherson05-qubit}.

A wide class of non-Gaussian states can be obtained by application of non-Gaussian operations on
Gaussian states, e.g., photon counting on squeezed states \cite{heralded}. The corresponding
non-Gaussian Wigner functions can, in turn, be expressed in terms of simple mathematical operations on
Gaussian functions. This implies that results explicitly derived for Gaussian states can be used as
generating functions for quantities of relevance also for non-Gaussian states. Closer to the spirit of
the present paper, we shall, however, give a few examples, where we explicitly apply the method of
Sec. III, \textit{i.e.}, we assume that the Wigner functions are known for the input states to the
protocol, and we carry transformations on the joint Wigner function of the entire physical system.

We shall focus on teleportation by use of a two-mode squeezed state, and we shall apply the same
protocol as above, but the input states will be number states and randomly displaced number states.

The entanglement channel of modes 1 and 2 is described by a covariance matrix $\gamma_{12}$, and hence
by the two-mode Wigner function,
\begin{equation}
W_{12}(x_1,p_1,x_2,p_2)=
\frac{1}{\pi\sqrt{\textrm{det}(\gamma_{12})}}\exp(-\chi^T\gamma_{12}^{-1}\chi),
\end{equation}
where $\chi^T = (x_1,p_1,x_2,p_2)$. We assume the covariance matrix in Eq.(10), with the allowed
values of the parameters $n$ and $k$, listed in Sec. IV.B.

The $N=1$ Fock state input Wigner functions is given by
\begin{equation}
W_{N=1}(x_3,p_3)=\frac{1}{\pi}(2x_3^2+2p_3^2-1)\exp(-x_3^2-p_3^2). \end{equation}
The Wigner functions
for the Fock states are all products of a polynomium in the arguments and a Gaussian. The beam
splitter operations, the evaluation at the arguments measured, and the integration over conjugate and
unmeasured variables, specified for the teleportation protocol in Sec. III, all preserve this
mathematical form of the Wigner function, and it is hence possible to obtain analytically the outcome
of the protocol and its fidelity. We shall now summarize the results of this analysis.

We have evaluated the output state for different degrees of entanglement of the teleportation channel.
For $N=0$ we reproduce the results of Sec. IV.B with $v_c=0$, and for higher $N$ we compare our
results with Ref.\cite{Caves04}. Our parameters $n$ and $k$ are equivalent to \textsf{c} and
\textsf{s} in Ref.\cite{Caves04}, and in the expression for the fidelity, our parameter $\Delta\equiv
n-k$ is equivalent to $t/2$ in the notation of \cite{Caves04}. (The expression $t\equiv
2/(\textsf{c}+\textsf{s})$ in \cite{Caves04} only applies for the pure state case, and should in the
general case be replaced by $2(\textsf{c}-\textsf{s})$ for the ensuing results to be correct). With
these modifications, we reproduce the expression for the fidelities in \cite{Caves04}, and in
particular the result
\begin{equation}\label{f-caves}
\ F_{N=1}=\frac{1+\Delta^2}{(1+\Delta)^3}.
\end{equation}

The analysis in \cite{Caves04} assumes a unit feedback. Applying instead a variable feedback gain $g$
as in the previous sections, we are able to optimize the teleportation protocols also for non-Gaussian
states: For the $N=1$ Fock state, we find that for strong entanglement (small $\Delta$) unit gain is
favored, but for weaker entanglement, it is advantageous to reduce the gain factor continuously to the
value $g=1/\sqrt{2}$ for $n=1$ and $k=0$. This is summarized by the numbers\\
\noindent  $\{($n$,\ F_{N=1,g=1},\ F_{N=1,g_\textrm{opt}})\}=\{(8, 0.8364, 0.8524),
(4,0.7098,0.7346)$, $(2,0.5258,0.5602)$,$(1,1/4=0.25,8/27=0.2963)\}$ for pure state channels with
$n^2=k^2+1$.

Teleportation of a known quantum state can in principle be replaced by a local production of the given
state with much higher fidelity. Let us therefore proceed with teleportation of \textit{unknown}
states, and let us begin with the teleportation of a Fock state, taken from an exponential
distribution of Fock states, $p_N\propto \exp(-N/\overline{N})$, with mean value $\langle
N\rangle=\overline{N}$. This distribution can also be written $p_N=(1-\lambda)\lambda^N$ with
$\lambda=\exp(-1/\overline{N})$. The average fidelity, according to our Eq.(4) is the mean value of
the fidelities weighted with the probability distribution for the input states. In \cite{Caves04}, the
Fock state teleportation fidelities are derived from a generating function, which is, apart from a
factor, precisely this mean value. We therefore readily obtain the mean fidelity as function of the
channel EPR variance and the parameter $\lambda$
\begin{equation}\label{lambda-form}
F(\lambda,\Delta)=\frac{1-\lambda}{\sqrt{(1+\Delta)^2-2\lambda(1+\Delta^2)+\lambda^2(1-\Delta)^2}}.
\end{equation}
The exponential distribution of $N$-values is equivalent to a Gaussian distribution of amplitudes,
hence the ensemble of Fock states has the same density matrix as the ensemble of displaced vacuum
states treated in Sec.IV, but since we are dealing with the state-to-state teleportation fidelity, the
results are very different. In particular, we found in (14), that when the variance of the
distribution of coherent input states diverges, the fidelity approaches $F=1/(1+\Delta)$, whereas
(\ref{lambda-form}) vanishes for fixed $\Delta$ and $\overline{N}=(v_c-1)/2\ \rightarrow \infty$. It
requires a very strongly entangled channel, $\Delta < 1/\overline{N}$, to reliably handle the
difference between highly excited Fock states.

In contrast, we have also implemented the scenario in which the input to the teleportation channel is
the $N=1$ Fock state displaced by an unknown amount, similar to the displacements of the vacuum
($N=0$) state in Sec. IV. With unit gain, in this case, we find that the result (\ref{f-caves}) holds
irrespectively of the variance $v_c$ of the distribution of displacements.

\section{Conclusion}

In conclusion, we have presented a theory to determine the fidelity of a general quantum state
transformation on an unknown quantum system. The result of this analysis is that, as long as the
protocol has been definitely determined in terms of the actions on the system conditioned on the
measurement outcomes, one can compute the fidelity as a simple weighted average of the state-to-state
fidelities over the incoming set of states.

We have introduced a formalism which incorporates the preparation of the input states, and showed that
our use of a fictitious system in a mixed state which is correlated with the input state to the
protocol may indeed be convenient for practical calculations. We demonstrated this last point in the
case of Gaussian transformations of Gaussian states, where we showed that the covariance matrix
formalism readily identifies the optimum performance and provides simple analytical results for the
fidelity of teleportation and quantum memories. The optimal use of non-trivial feed-back gain in these
protocols was a particularly interesting result brought out clearly by the analysis. Finally we showed
that more general states can also be handled by their appropriate Wigner functions.

\begin{acknowledgments}
 We thank Jacob Sherson, Uffe V. Poulsen and Eugene Polzik for useful discussion. LBM was supported by the Danish
 Natural Science Research Council (Grant No. 21-03-0163).
 \end{acknowledgments}
%\bibliography{bibfile}
%\bibliographystyle{apsrev}

\end{document}